\documentclass[conference]{IEEEtran}

\usepackage{graphicx}
\usepackage{epsf}
\usepackage{amsmath, amsfonts}
\usepackage{latexsym}
\usepackage{subfigure}
\usepackage{multirow}
\usepackage{multicol}
\usepackage[table]{xcolor}
\usepackage[hidelinks]{hyperref}
\usepackage{algorithm2e}

\begin{document}

\title{\Huge Danger Aware Vehicular Networking}

\author{
Tsigigenet Dessalgn and Seungmo Kim\\
		
		\IEEEauthorblockA{Department of Electrical and Computer Engineering\\
			Georgia Southern University\\
			Statesboro, USA\\
			\{td04611, seungmokim\}@georgiasouthern.edu}
}
	
\maketitle		
\begin{abstract}
IEEE 802.11p is one of the key technologies that enable Dedicated Short-Range Communications (DSRC) in intelligent transportation system (ITS) for safety on the road. The main challenge in vehicular communication is the large amount of data to be processed. As vehicle density and velocity increases, the data to be transmitted also increases. We proposed a protocol that reduces the number of messages transmitted at a vehicle according to the level of danger that the vehicle experiences. The proposed protocol measures inter-vehicle distance, as the representative of the danger of a vehicle, to determine the priority for transmission. Our results show that this prioritization of transmissions directly reduces the number of transmitters at a time, and hence results in higher performance in terms of key metrics--i.e., PDR, throughput, delay, probabilities of channel busy and collision.
\end{abstract}

\vspace{0.2 in}


\section{Introduction}
According to the World Health Organization (WHO), approximately 1.2 million people are killed each year in road crashes and as many as 50 million are injured \cite{cps_1}. Vehicle-to-everything (V2X) communications \cite{cps_2} keep key potentials in enhancement of the intelligent transportation system (ITS), especially in applications for traffic safety. However, the concurrent V2X communications techniques must address (i) spectrum contention \cite{cps_3}-\cite{cps_6} and (ii) latency \cite{cps_7}\cite{cps_8}.

This paper, discusses a V2X communications framework that is adaptive to the accident. In order to address contention among vehicles for access to the channels, this paper proposes to (i) prioritize a message according to the level of accident risk and (ii) filter a message from transmission, which relieves congestion of the communications network traffic.

\subsection{Related Work}
Radio communications have been acting a significant role so far in keeping traffic safety in ITS. Two main approaches to establishment of a radio communications system for road infrastructure are DSRC and cellular-V2X (C-V2X).

C-V2X is a concept involving the use of cellular communications for vehicle connectivity use cases and applications. Many of these use cases are safety-related, though there are also mobility- and environment-related use case opportunities as well, and the breadth of use cases is increasing. However, C-V2X is still questionable about its capability of delivering the latency requirement \cite{cps_7}-\cite{cps_8}.

Integration of the two technologies was discussed to complement each other's drawbacks \cite{cps_9}-\cite{cps_13}. Yet none of the two technologies can provide a clear guideline in addressing higher \textit{spectrum contention} that is expected in future V2X environments. Largely increased number of vehicles and infrastructure that will be equipped with communications functionality \cite{cps_14} will make integration of the two different types of communications networks remarkably complicated. The problem is that the higher contention will lead to higher probabilities of \textit{disconnection} \cite{globecom18}. For instance, such vehicles at a junction in a large city at a rush hour can experience disconnection due to congestion of too many messages exchanged over the air, which increases the probability of a crash.

Adjusting the contention window (CW) size was highly used by different researchers to improve the performance of a network.  In vehicular communication, various mechanisms for changing size of CW have been developed. Though, changing the CW size may not always help to improve the performance of the network \cite{wu2018improving}. Another way, parameters mentioned on the dynamic control backoff algorithm (DCBTA) model proved that, the number of transmitter stations have a direct impact on the performance of the system. On the other hand, distance based routing protocol's performance is better for traffic load environment of vehicular ad-hoc networks (VANETs) \cite{ramakrishna2012dbr}. Also, in an ad-hoc network, reduction of the length of a header can be a solution that is worth considering \cite{grasnet}. However, a V2X network must be capable of operating without a `gateway,' which was suggested in \cite{grasnet}.

Safety messages are broadcast and periodic \cite{dissertation}. Using these characteristics of safety message a reverse back off mechanism was modeled. The method uses the expiration of periodic safety messages in order to decide the value of CW. The method helped to increase the reception probability. However, it considers limited lifetime of cooperative awareness message (CAM). A CAM can be transmitted only when back off time is zero. Moreover, there is a probability of the message to expire before MAC transmits them \cite{cps_5}.

Stochastic geometry has been adopted for analysis of the networking behaviors in a DSRC system \cite{arxiv19}-\cite{access19}. Considering the spatial dynamicity of vehicles, the sptial modeling is of paramount importance in evaluation of a network's performance. This importance motivated the main idea of this paper: ``distance'' as the key metric according to which the backoff time is allocated.

\subsection{Contributions}
Therefore, we proposed an algorithm which is used for vehicular network and the algorithm focuses on reduction of the number of transmitters at a time. Practical V2X environment needs a network system with better performance and low latency due to properties such as mobility, vehicle density and high spectrum contention properties of practical V2X environments. The algorithm considers saturated condition. Based on saturated condition of the vehicular network, DCBTA model is used for comparison.

As a solution for the spectrum contention, this paper proposes a V2X communications scheme where a vehicle takes the opportunity for a transmission according to the probability that it runs into a crash.

The technical contributions are two-fold:
\begin{enumerate}
    \item This is the first work that proposes resource allocation according to the danger.
    \item In order to measure the danger, this paper uses the `inter-vehicle distance' as the metric in setting the threshold.
\end{enumerate}

\vspace{0.2 in}

\section{System Model}
Distributed coordination function (DCF) is defined as the basic access mechanism of IEEE 802.11 medium access control (MAC) \cite{cheng2014novel}. DCF is regularly used in vehicular communication in a contention-based manner \cite{wu2018improving}. The main purpose of this paper is to evaluate and enhance the performance of a system in a vehicular communication using the well-known parameters such as throughput $(S)$ and packet delivery rate $(PDR)$. We first obtain the stationary probability $(\tau)$ by studying the behavior of a single station. By using the bi-dimensional Markov chain model we were able to calculate throughput $(S)$ and packet delivery rate $(PDR)$ from the the stationary probability $(\tau)$. Thus, the two-dimensional stochastic process ($s(t)$, $b(t)$) is used to describe the behavior of a single station. Where $b(t)$ and $s(t)$ represent the backoff time counter for a given station and the backoff stage of the station at a time respectively \cite{bianchi2000performance}.

In a random 802.11 DCF, a station first listens to the activity of the channel in order to transmit a new packet. If the channel is idle for a period of Distributed Inter frame Space (DIFS) time, the station transmits. However, if the channel is busy during the DIFS time, the station generates random backoff time before transmitting the packet. In this case, the time value is represented by contention window (CW) size and the transmitter station selects a random backoff time counter using the following equation:

\begin{equation}\label{eq1}
\text{Backoff Counter} = Random()\times T_{slot}
\end{equation}
where $T_{slot}$ denotes slot time; $Random$ is a random integer within $[0, CW]$ with $CW \in [W_{min}, W_{max}]$. The values for $W_{min}$ and $W_{max}$ denoted in last version of the standard \cite{IEEE Standard } in the PHY-specific section. The backoff counter starts to decrease if the channel is detected idle. According to that, the backoff counter will freeze when the channel is busy. On the other hand, When the backoff counter reaches zero the station is allowed to start transmitting the packets. 

In this paper, a two-dimensional Markov chain analysis model, which is derived as a stochastic model, is used to evaluate the performance of IEEE 802.11p DCF. Although, we make different assumptions for the evaluation of vehicle to vehicle communication. In a model that was introduced in \cite{bianchi2000performance}, the probability of packet transmission $(\tau)$ was assumed to be dependent on the collision probability $(P_{c})$. However, \cite{alkadeki2014estimation} model has introduced a new probability called the busy probability $(P_{b})$ using the stationary distribution $b_{i,k}$ where $i\in (0,m)$ in which, the maximum backoff stage is represented by $m$ and $k$ represents the backoff time counter. In fact, when $k=0$ a transmission has occurred. Meanwhile, the busy probability and the collision probability are considered as two independent process in MAC transmission. By definition, a collision occurs when multiple vehicles (transmitter stations) try to access the transmission medium simultaneously and the busy probability $(P_{b})$ is, referring to the probability of the channel being busy.

The Markov chain model uses five transition probability sates for IEEE 802.11 DCF transmission system. Idle state, successful transmission state, busy state, collision state at initial stage ($i$) and collision state at maximum stage ($m$) are the five transition probabilities used for determining the stationary distribution $b_{i,k}$. They are described using the following mathematical equations \cite{bianchi2000performance}.
\begin{align}\label{eq2}
P({i,k| i,k+1})=\frac{1-P_{b}}{W_{i}} , k\in  (0,{W_{i-2}}), i\in (0,m)
\end{align}
\begin{align}\label{eq3}
P({0,k| i,0})=\frac{(1-P_{c})}{W_{0}} , k \in (0,{W_{0-1}}), i\in (1,m)
\end{align}
\begin{align}\label{eq4}
P({i,k| i,k})=\frac{P_{b}}{W_{i}}, k \in (0,{W_{i-1}}), i\in (0,m)
\end{align}
\vspace{0.01 in}
\begin{align}\label{eq5}
P({i,k|,i-1,0}) =\frac{P_{c}}{W_{i}}, k \in (0,{W_{i-1}}), i\in (1,m) 
\end{align}
\begin{align}\label{eq6}
P({m,k|,m,0}) =\frac{P_{c}}{W_{m}}, k \in (0,{W_{m-1}}), i = m 
\end{align}
The decrement of the backoff time counter is represented in Eq. (\ref{eq1}). The backoff time counter after a successful transmission always starts with the backoff stage of `0' as expressed in Eq. (\ref{eq2}). If there is unsuccessful transmission at backoff stage $i$, the backoff time will be chosen randomly among (0, $CW_{min}$) \cite{bianchi2000performance}.

The discrete-time Markov chain model will be used to analyze the two-dimensional stochastic process $s(t)$,  $b(t)$. Assume $b_{i,k}=\lim_{t\to\infty} P\big\{s(t)=i, b(t)=k\big\}, i\in (0,m), k\in (0, {W_{0-1})}$ as the stationary distribution and the  busy probability $(P_{b})$ is introduced from this stationary distribution.  

On the other hand, to find $\tau$, the basic parameter for throughput $(S)$ and packet delivery rate $(PDR)$ computation, we have used the discrete-time Markov chain process. In this case, it is expressed by $(i,k)$, using this we can have various states for a single station. In this case, $b_{i,k}$, $b_{m,0}$, $b_{m,k}$, $b_{0,k}$, $b_{0,0}$ and $b_{i,0}$ are the different states used to describe a single station and they are used to compute $\tau$ \cite{alkadeki2014estimation}. Technically, the stationary probability $b_{i,k}$ can be used to represent the other states. 
\begin{equation}\label{eq7}
b_{0,k}=b_{0,0}\frac{1}{P_{b}-W_{0}}, \forall k\in (1,W_{0}-1) 
\end{equation}
\begin{equation}\label{eq8}
b_{i,0}={P_{c}^{i}b_{0,0}, \forall i\in (1,m-1)}
\end{equation}
\begin{equation}\label{eq9}
b_{i,k}=b_{0,0}\frac{P_{c}^{i}}{1-\frac{P_{b}}{W_{0}}}(1-\frac{k}{W_{i}}), \forall i\in (1,m-1)
\end{equation}
\begin{equation}\label{eq10}
b_{m,0}=\frac{P_{m}}{{1-P{c}}^{i}}b_{0,0},
\end{equation}
\begin{equation}\label{eq11}
b_{m,k}=b_{0,0}(1-\frac{k}{W_{m}})\frac{1}{1-\frac{P_{b}}{W_{m}}}\frac{P_{c}^{m}}{{1-P{c}}}, \forall k\in (1,W_{m}-1) 
\end{equation}
Hence, these different probability states should give a total probability mass $1$ we used normalization condition to find the remaining probability state $b_{0,0}$. Notice that $b_{0,0}$ is the unknown quantity. Therefore, to find for $b_{0,0}$ we used the normalization condition as follow as:
\begin{equation}\label{eq32}
1=\sum_{i=0}^{m} \sum_{k=0}^{w_{i}-1}b_{i,k}
\end{equation}
once we have $b_{0,0}$, we can plug in to Eq.(7)-Eq.(11) to find the other probability states. In addition, in previous discussion it was mentioned that for a transmission to occur $k=0$. Thus, the probability that a station can transmit can be expressed using: 
\begin{equation}\label{eq14}
\tau=\sum_{i=0}^{m}b_{0,0}
\end{equation}
By having this, the probability of at least one  station transmitting in a slot can be also calculated as:
\begin{equation}\label{eq12}
P_{tr}=1-(1-\tau)^n
\end{equation}
Similarly, the  probability that a packet transmission is successful can be expressed as
\begin{equation}\label{eq13}
P_{su}=\frac{n\times \tau\times (1-\tau)^{n-1}}{1-(1-\tau)^n}
\end{equation}
where $n$ is the number of contending stations at a time.

\vspace{0.2 in}

\section{Proposed Protocol}
In this paper, we consider a two directional road with one lane. We assumed each vehicle knows everything about the road. Including, the road map, location and distance information of other vehicles. Following the typical parameter setup for an IEEE 802.11p system \cite{cps_2}, the transmission range is assumed to be 1,000 m. Hidden nodes are not considered and each of the vehicle's packet load size is similar (1,023 bytes). Moreover, the distance in which packets can be successfully sent/received is known as transmission range $(R)$ \cite{yao2013performance}. We assumed 50 randomly distributed vehicles  to be on the lane and each vehicle is aware of the distance from a vehicle with its immediate preceding and subsequent vehicles. The decision for the number of transmitter's at a specific time, depends on the distance information. In a two directional road with one lane that we assumed, if we have one vehicle at a point $(x_{1},y_{1})$ and another vehicle at a point $(x_{2},y_{2})$. Thus, the distance between these two vehicles is given by the following normal distance equation:

\begin{equation} \label{eq15}
D =\sqrt{(x_{2}-x_{1})^2+(y_{2}-y_{1})^2}
\end{equation}
				
Having the distance information between each vehicle, priority is given for those vehicles which are within the threshold value. In this case, the threshold distance is a distance in which vehicles are allowed to transmit. Threshold value of 300 m, 500 m and 700 m have been used in our simulation. 

Mainly, priority of message depends upon message urgency and dissemination distance \cite{suthaputchakun2012priority}. 
Assuming 50 vehicles are on the road at a specific time all vehicles will not be considered as a transmitter station unless they are within the threshold distance. Thus, the approach we propose reduces the number of transmitters. Eventually, each transmitter will be assigned to transmit based on their specific distance information. The proposed protocol is simulated using MATLAB. The algorithm we used for the simulation results will be discussed later.

In section I, we define the concept of CAM message and it was observed that, while waiting for the backoff time to be zero, there is a probability that an important message will expire before it is transmitted. However, in our proposed algorithm the transmitter vehicle is not necessarily waiting for the backoff time to be zero. Instead, it will transmit the message to the vehicle in front or behind if it is located within the threshold distance. Moreover, The decision to transmit data is always dependent on how far the vehicle is located to the vehicle it needs to communicate. It is due to safety messages are more important to the neighbor vehicles and that is the reason why the decision for the value of CW should depend on the distance between the vehicles.

\begin{figure}[t]
\centering
\includegraphics[width=\linewidth]{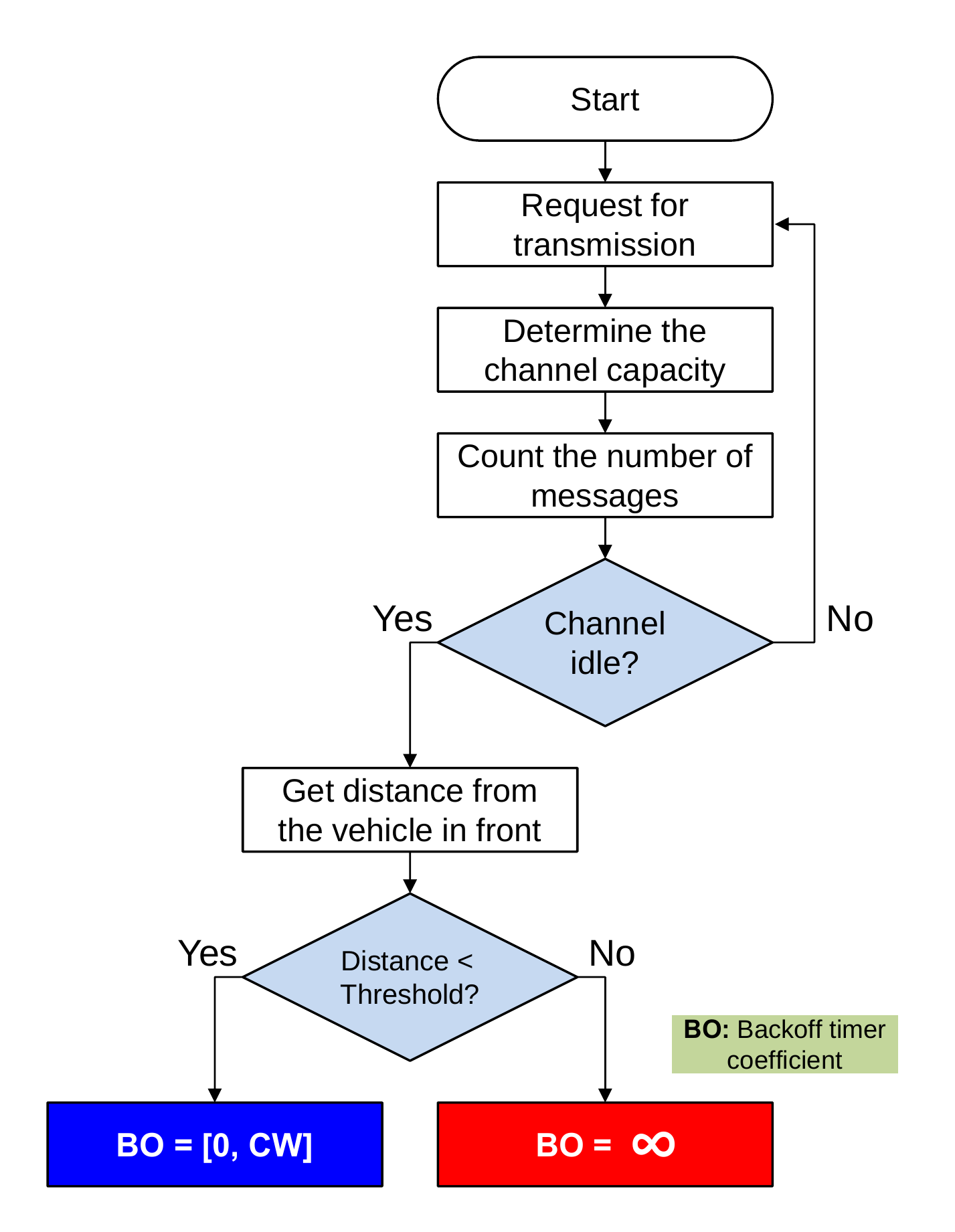}
\caption{Flowchart for the proposed algorithm}
\label{fig_flowchart}
\end{figure}

As it can be seen from the flowchart, the following steps have been used to build the algorithm:
\begin{itemize}
\item \textbf{Step 1:} 50 vehicles are uniformly randomly distributed within a linear space that is 1,000 m long.
\item \textbf{Step 2:} The distance between each vehicle is calculated.
\item \textbf{Step 3:} A threshold value is applied. (300 m, 500 m or 700 m was used for simulations.)
\item \textbf{Step 4:} The danger of a vehicle is measured according to the inter-vehicle distance in comparison to the threshold value.
\item \textbf{Step 5:} If the distance is less than threshold, the vehicle is granted the opportunity for a transmission.
\end{itemize}

\vspace{0.2 in}

\section{Performance Analysis}\label{sec_analysis}
\subsection{Packet Delivery Rate and Throughput}
The packet delivery rate (PDR), which is the rate at which data are successfully delivered to a destination is compared to the amount of data been sent out, and throughput are parameters used to determine the performance of a system and it is clear that as the number of transmission decreases, the value of PDR and throughput also increases. Besides, when the transmission channel is idle after the DIFS period the backoff timer continues to decrease until zero and packet from the transmitter station would be transmitted. In case, the transmission medium becomes busy while the backoff timer is decremented the backoff timer will freeze until the channel becomes idle again. The period that the back of timer will freeze is called busy probability. Having the busy probability and the collision probability we are able to calculate the packet transmission probability ($\tau $) \cite{alkadeki2015performance}. Throughput $(S)$, a widely used parameter to determine the performance of a system, can be computed using the following expression \cite{bianchi1998ieee}
\begin{equation}\label{eq16}
S=\frac{P_{su}P_{tr}E[P]}{(1-P_{tr})+P_{tr}P_{su}T_{s}+P_{tr}(1-P{su})T_{c}}
\end{equation}
where $E[P]$, $P_{tr}$, $P_{su}$, $T_{c}$, $T_{s}$ and $\delta$  represents the average packet length, transmission probability, successful packet transmission probability, collision time, successful transmission time and propagation delay, respectively. 
\begin{equation}\label{eq17}
T_{s}=H+E[P]+SIFS+\delta+ACK+DIFS+\delta
\end{equation}
\begin{equation}\label{eq18}
T_{c}=H+E[P]+DIFS+\delta
\end{equation}

\subsection{Packet Delay}
The time taken by a packet to travel from source to destination is called delay. Delay is denoted as a sequence of intervals of empty delay time $(D_{emp})$, successful delay time $(D_{suc})$, busy delay time $(D_{bus})$ and collision delay time $(D_{col})$ \cite{ivanov2011delay}\cite{alkadeki2014estimation}. Moreover, the behavior of the proposed model  is based on two probabilities. The probability of a vehicle trying a transmission  ($\tau_{tr}$) and  a probability of simultaneous transmission  ($\tau_{nb}$).  $P_{emp} $, $P_{bus}$, $P_{suc}$, $P_{own}$ and  $P_{col}$, which are probability of  channel being idle, probability  of channel busy,  probability of successful transmission, probability of a station attempting transmission and probability of collision, have been considered for analytic determination of MAC layer packet delay distribution. The value for $\tau_{tr}$ and $\tau_{nb}$  is considered to have the same  as $\tau$ value.

\begin{equation}\label{eq19}
{P_{emp}=(1-\tau_{tr})(1-\tau_{nb})^{n-1}}\\
\end{equation}
\begin{equation}\label{eq20}
{P_{suc}=(n-1)(\tau_{nb})(1-\tau_{tr})(1-\tau_{nb})^{n-2}}\\
\end{equation}
\begin{equation}\label{eq21}
{P_{own}=\tau_{tr}(1-\tau_{nb})^{n-1}}\\
\end{equation}\begin{equation}\label{eq22}
{P_{col}=\tau_{tr}({n-1})(\tau_{nb})(1-\tau_{nb})^{n-2}}	\\
\end{equation}
\begin{equation}\label{eq23}
{P_{bus}=1-P_{emp}-P_{own}-P_{suc}-P_{col}}\\
\end{equation}

Thus, the MAC layer delay can be presented as a terminating renewal process, which terminates with probability of each successful transmission and can be presented as a sequence as $S_n = T_{1}+T_{2}+T_{3}+.......T_{n}+D_{suc}$.

Our proposed algorithm have been able to reduce the number of transmissions, which  helps to improve the throughput and PDR of the system. In addition,  it also helps to reduce the probability of channel being busy and the probability of collision, which is shown in Fig. 5 and Fig. 6.  When we have a reduction on all the mentioned parameters and the backoff time, the total time delay is decreased. This proved that the proposed algorithm performs better compared to the base line.   					
The total delay $(T_{td})$ time can be calculated from the total sum of  transmission time $(T_{tt})$, total time delay in the collision $(T_{tc})$, idle time $(T_{emp})$ and average backoff time $(CW^{*})$
\begin{equation}\label{eq24}
T_{td}= (T_{tt}) +(T_{tc})+(CW^{*})+T_{emp}\\  
\end{equation}
\begin{equation}\label{eq25}
T_{tt}= T_{tsp}\times N_{transmission}\\
\end{equation}
\begin{equation}\label{eq26}
T_{tc}=T_{tsc}\times N_{collision}\\   
\end{equation}
\begin{equation}\label{eq27}
N_{transmission}= P_{tr}\times N_{transmitter}\\
\end{equation}
\begin{equation}\label{eq28}
N_{collision}= P_{col}\times N_{transmitter}\\
\end{equation}
where $T_{tsp}$  is transmission time of single packet, $T_{tsc}$ is total time in collision, $T_{emp}$ is idle time and $CW^{*}$ is back off time.
\begin{equation}\label{eq29}
 T_{tsp}= RTS+CTS+3SIFS + Data + ACK + DIFS\\
\end{equation}
\begin{equation}\label{eq30}
T_{tsc}=RTS + DIFS
\end{equation}	
The average back of time is $CW^*$ formulated as
\begin{equation}\label{eq31}
CW^{*}=\frac{CW_{min}\times T_{slot}}{2}
\end{equation}
where $T_{slot}$ is a slot time and $CW_{min}$ is minimum backoff window size \cite{park2014coexistence}\cite{vinel20123gpp}. As mentioned in Section \ref{sec_analysis}, the equation for the probability of transmission $(P_{tr})$ and probability of collision ($P_{col}$) is stated on Eq. (14) and Eq. (23). Average number of collision can be calculated by using $(P_{col})$ and average number of transmitters, the same logic can be used for average  number of transmission computation. Finally, by using Eq. (25), Eq. (32) and Eq. (1) the total time delay is calculated. As it can be seen from the below figure, the proposed algorithm reduces the total time delay regarding the threshold distance .

\begin{figure}[t]
\centering
\includegraphics[width=\linewidth]{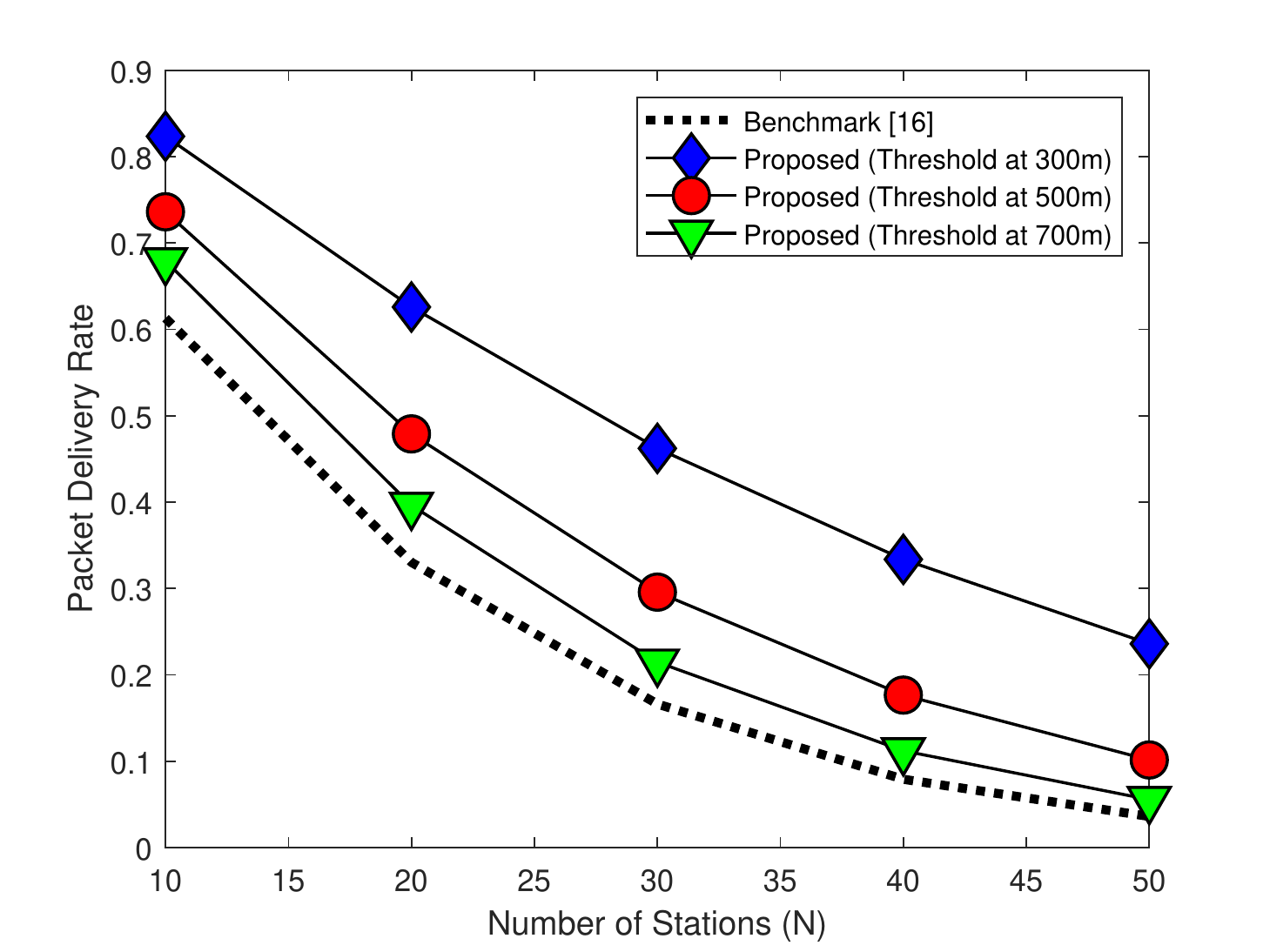}
\caption{Packet delivery rate}
\label{fig_pdr}
\end{figure}

\begin{figure}[t]
\centering
\includegraphics[width=\linewidth]{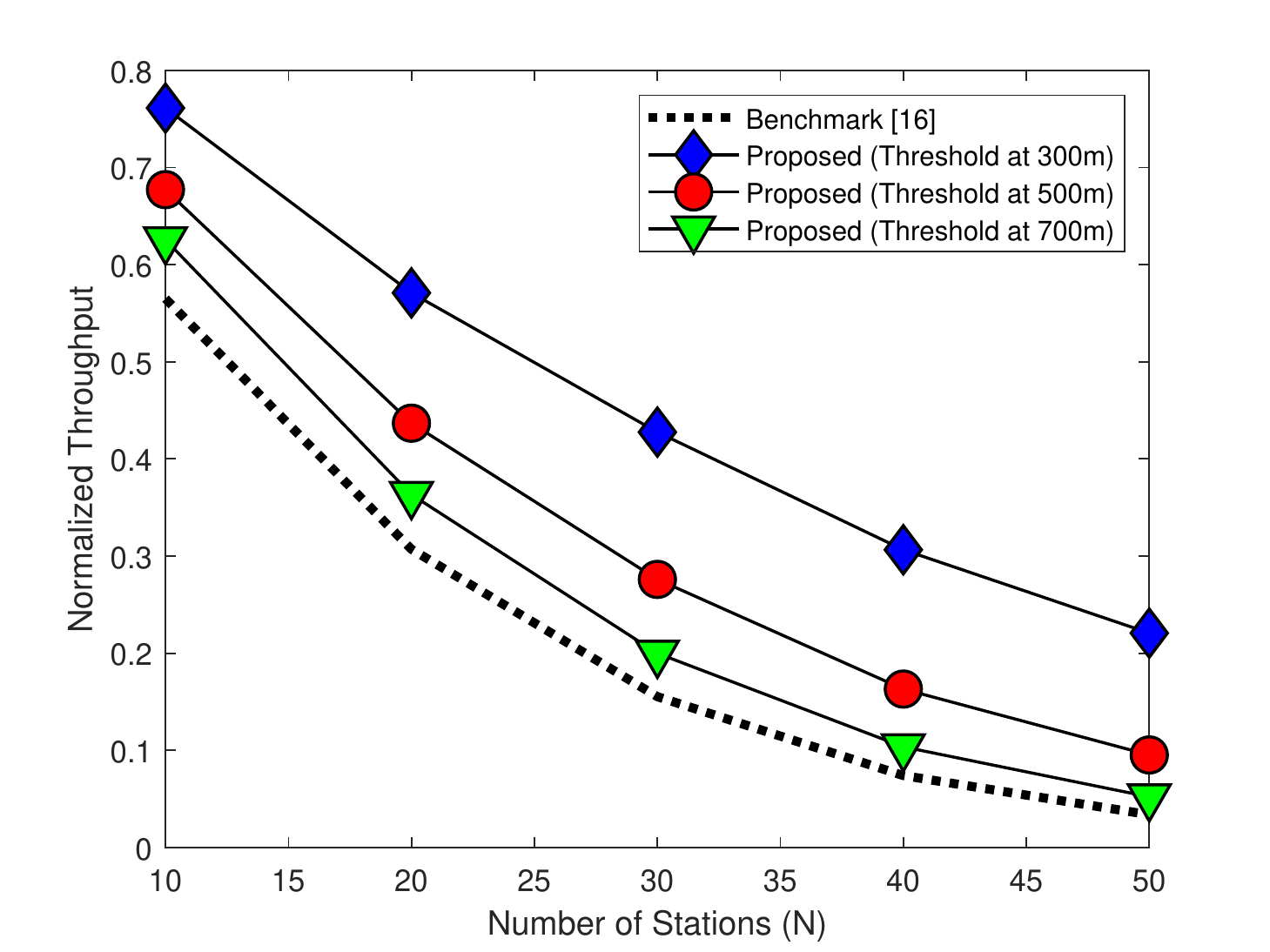}
\caption{Throughput}
\label{fig_throughput}
\end{figure}

\begin{figure}[t]
\includegraphics[width=\linewidth]{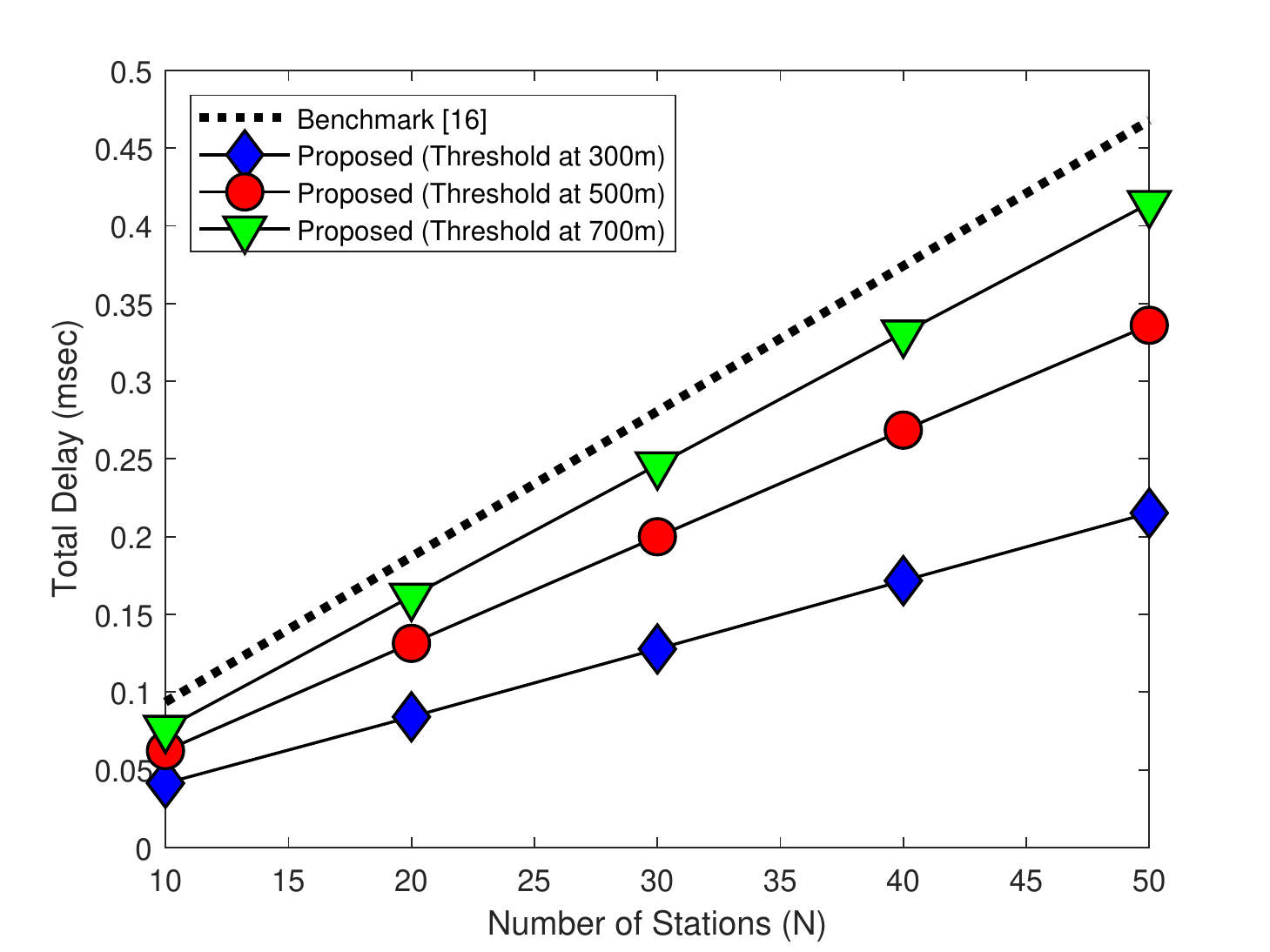}
\caption{Total delay}
\label{fig_delay}
\end{figure}

\vspace{0.2 in}

\section{Numerical Results}
We validate our result for the proposed algorithm. In our simulation we consider the number of vehicle on the road to be 50 in a road length of 1,000 m. The effective communication range is the as same as the road length. We assume the distance among vehicles to be in a random distribution and each vehicle can send message each time.

\begin{table}[t]
\caption{Parameters for simulations}
\centering
	\begin{tabular}{ |p{3cm}|p{3cm}|}
		\hline
		\textbf{Parameter} & \textbf{Value}\\
		\hline\hline
		DIFS& 64 $\mu$s\\
		SFIS& 32 $\mu$s\\
		Payload length&1023 bytes\\
		Propagation delay $(\delta)$ & 1 $\mu$s\\
		Slot time & 13 $\mu$s\\
		$CW_{min}$& 7\\
		\hline
	\end{tabular}
\end{table}

Fig. 2 shows PDR as one of the key metrics measuring the performance of the proposed algorithm. The curves with blue, red and green dots corresponds to the packet delivery rate at 300 m, 500 m and 700 m. It is observed that the scheme we proposed have a better performance than the benchmark and when we decrease the threshold distance the performance from the benchmark is getting better and better. In addition to the PDR curve another parameter to show the performance of the proposed system is throughput. Thus, throughput is presented in Fig. 3. Obviously, the proposed algorithm performs with higher throughput than the benchmark. The explanation of the curve with blue, green and red dots is similar to the PDR curve.

Fig. 4 shows the total delay for the proposed algorithm. In previous discussion, it is stated that the total delay time is the sum of transmission time, total collision time and average backoff time. Using the algorithm we propose it is possible to reduce the effect of each component's of the total delay time. As we decease number of the contending stations, the packet size to be transmitted also decreases and a reduced packet size can be transmitted easily. In the mean time, the reduction in the number of transmitters will help to reduce the probability of collision, which is also explained in Fig. 6.

The smaller the number of packets received, the lower the probability of channel being busy. The channel busy probability is illustrated in Fig. 5. Hence, the number of transmitter at a threshold distance of 700 m is less than the one in 500 m and 300 m and the channel is less busy as we decrease the threshold distance. Furthermore, the backoff time before transmission can also be reduced with reduced number of transmitter and the backoff time has an effect on the probability of the channel being busy.

Fig. 6 illustrates the probability of collision. The achieved probability of collision is mush lower than the benchmark. Therefore, the probability of packets being collided during transmission is reduced in the proposed algorithm. The curve in Fig. 6 clearly shows as we decrease the threshold distance packets collision decreases.

\begin{figure}[t]
\centering
\includegraphics[width=\linewidth]{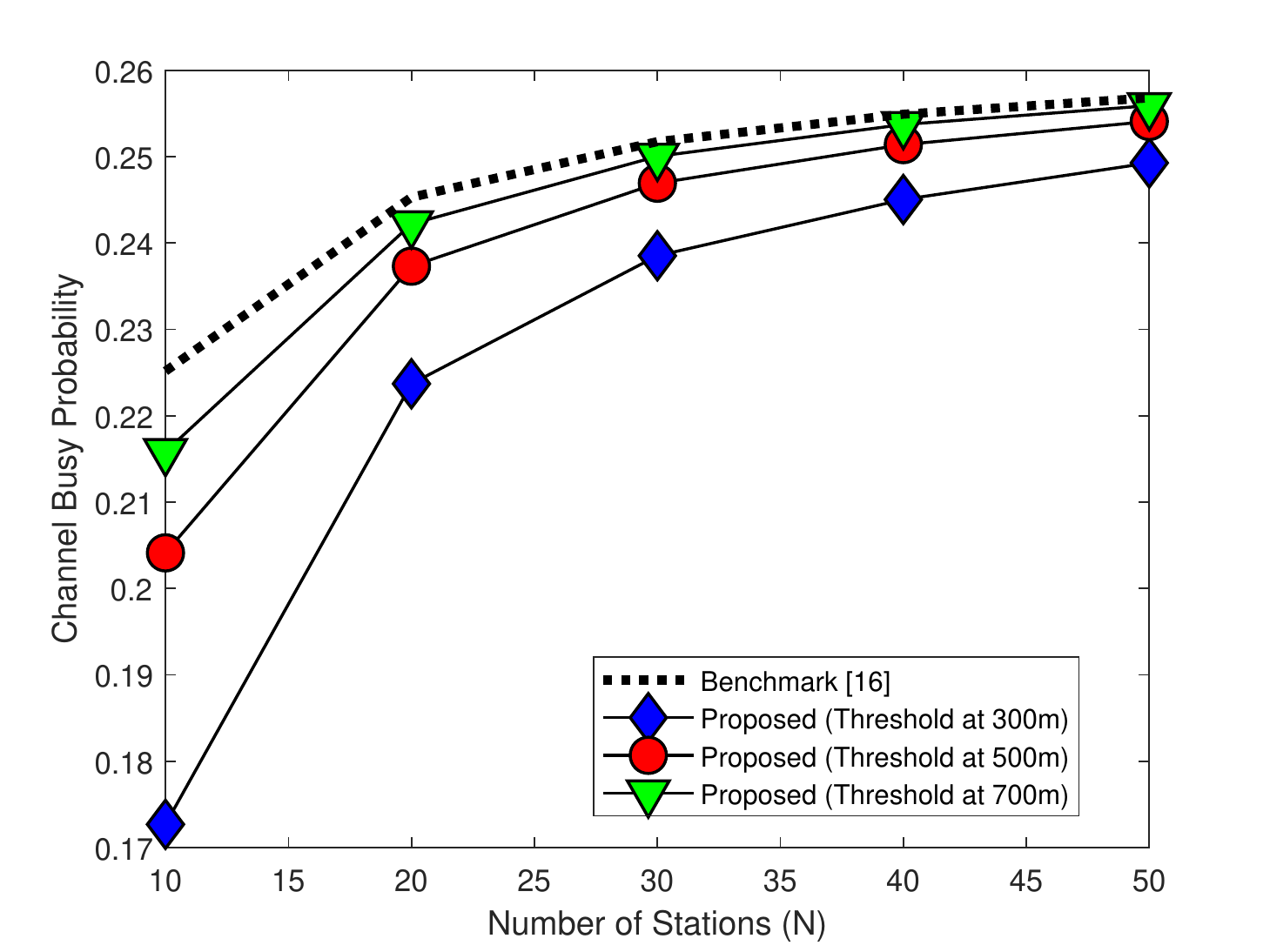}
\caption{Channel busy probability}
\label{fig:5}
\end{figure}

\begin{figure}[t]
\centering
\includegraphics[width=\linewidth]{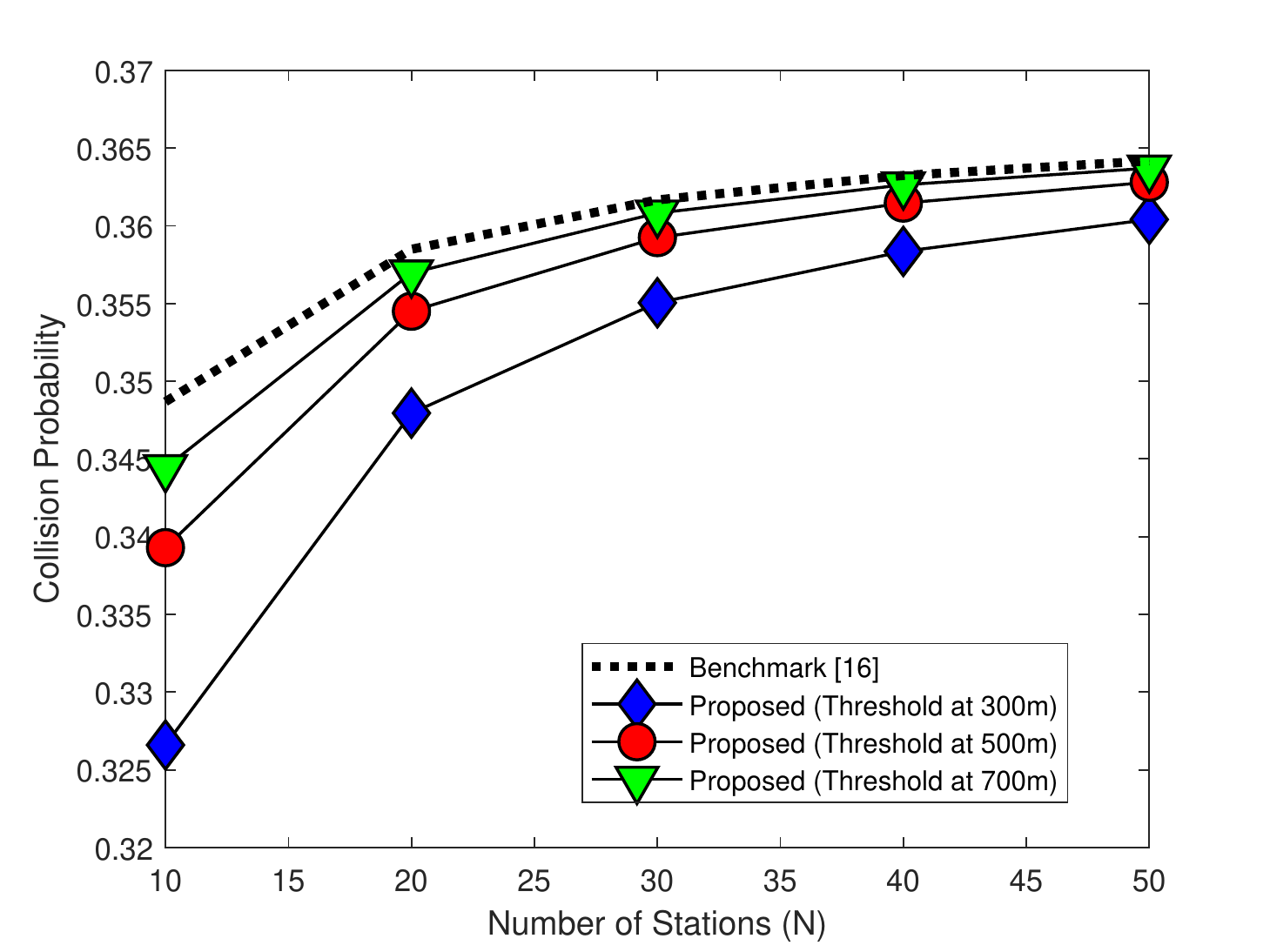}
\caption{Probability of collision}
\label{fig:6}
\end{figure}

\vspace{0.2 in}

\section{Conclusion and Future Work}
In this paper, we proposed an algorithm to improve the performance of a V2X communications network via danger-aware allocation of a transmission opportunity. A danger level has been measured by relying on the inter-vehicle distance. The proposed scheme takes a threshold value on the inter-vehicle distance in order to determine a transmission at a vehicle. Our simulation results showed that application of the proposed protocol could increase the performance of a V2X network, by allowing only the vehicles at higher danger to transmit. The performance was measured in various metrics--i.e., PDR, throughput, probability of collision, channel busy probability, and total delay.

We identify two specific tasks as the future work. First, we will improve the proposed protocol in such a way that `smoother' resource allocation is achieved. In other words, improving the current classification of `under' and `over' a threshold, the next version of our work will more finely classify the danger into a larger number of levels and allocate the backoff time constants according to the danger. Second, a danger level will be determined via integration of more parameters in addition to inter-vehicle distance. While the inter-vehicle distance is a key quantity that leads to various dangerous scenarios, a crash is attributed from a variety of other variables such as driver-related factors. We will develop a more rigorous danger indicator incorporating the various factors.

\vspace{0.2 in}



\end{document}